\documentclass[aps,prl,twocolumn,nobibnotes,showpacs,superscriptaddress]{revtex4-1} % for double-spaced preprint
\usepackage{graphicx} % needed for figures
\usepackage{dcolumn} % needed for some tables
\usepackage{bm} % for math

\usepackage{amssymb} % for math
\usepackage{amsmath} % for math
\usepackage[alsoload=hep]{siunitx} % nice SI units 
\usepackage{color}
\usepackage{xcolor}
\usepackage{hyperref}% add hypertext capabilities

\newcommand{\ee}{\ifmmode (e,e') \else $(e,e')$~\fi}
\newcommand{\Aee}{\ifmmode A(e,e') \else $A(e,e')$~\fi}
\newcommand{\eep}{\ifmmode (e,e'p) \else $(e,e'p)$~\fi}
\newcommand{\Aeep}{\ifmmode A(e,e'p) \else $A(e,e'p)$~\fi}
\newcommand{\pmiss}{\ifmmode p_{miss} \else $p_{miss}$\fi}
\newcommand{\emiss}{\ifmmode E_{miss} \else $E_{miss}$\fi}

\newcommand{\het}{\ifmmode ^3{\rm He} \else $^3$He\fi}
\newcommand{\trit}{\ifmmode ^3{\rm H} \else $^3$H\fi}
\newcommand{\hets}{\ifmmode ^3{\rm He} \else $^3$He \fi} %with space at the end (for normal text)
\newcommand{\trits}{\ifmmode ^3{\rm H} \else $^3$H  \fi} %with space at the end (for normal text)
\begin{document}

\title{Comparing proton momentum distributions in \bm{$A= 2$} and \bm{$3$} nuclei via $^2$H $^3$H and $^3$He \bm{$\eep$} measurements}

\author{R. Cruz-Torres} \affiliation{Massachusetts Institute of Technology, Cambridge, MA}
\author{S. Li} \affiliation{University of New Hampshire, Durham, NH}
\author{F. Hauenstein} \affiliation{Old Dominion University, Norfolk, VA}
\author{A. Schmidt} \affiliation{Massachusetts Institute of Technology, Cambridge, MA}
\author{D.~Nguyen} \affiliation{University of Virginia, Charlottesville, VA}

%A                                                                                                                                                            
\author{D.~Abrams} \affiliation{University of Virginia, Charlottesville, VA}
\author{H.~Albataineh} \affiliation{Texas A \& M University, Kingsville, TX}
\author{S.~Alsalmi } \affiliation{Kent State University, Kent, OH}
\author{D.~Androic} \affiliation{University of Zagreb, Zagreb, Croatia}
\author{K.~Aniol} \affiliation{California State University , Los Angeles, CA}
% ...

\collaboration{Jefferson Lab Hall A Tritium Collaboration}

\date{August 21, 2019}

% ==============================================================================================================
% Abstract
\begin{abstract}
We report the first measurement of the \eep reaction cross-section ratios for Helium-3 (\het), Tritium (\trit), and Deuterium ($d$). The measurement covered a missing
momentum range of $40 \le \pmiss \le \SI{550}{\mega\eVperc}$, at
large momentum transfer ($\langle Q^2 \rangle \approx 1.9$ (GeV/c)$^2$) and $x_B>1$,
which minimized contributions from non quasi-elastic (QE) reaction mechanisms.
The data is compared with plane-wave impulse approximation
(PWIA) calculations using realistic spectral functions and momentum distributions.
The measured and PWIA-calculated cross-section ratios for \het/$d$ and
\trit/$d$ extend to just above
the typical nucleon Fermi-momentum ($k_F \approx 250$ MeV/c) and differ from each other by $\sim 20\%$, while for 
\het/\trit{} they agree within the measurement accuracy of about 3\%.
At momenta above $k_F$, the measured \het/\trit{} ratios differ from the calculation by $20\% - 50\%$. 
Final state interaction (FSI) calculations using the generalized Eikonal Approximation indicate that
FSI should change the \het/\trit{} cross-section ratio for this measurement by less than 5\%.
If these calculations are correct, then the differences at large missing momenta between the
    \het/\trits experimental and calculated ratios  could be due to the underlying
    $NN$ interaction, and thus could provide new constraints on the previously
    loosely-constrained short-distance parts of the $NN$ interaction.
\end{abstract}

\maketitle

% ==============================================================================================================
% Body

% -------------------------------------------------------------------------------------------------------------
% Intro
Nuclear interaction models are a crucial starting point for modern
calculations of nuclear structure and reactions, as well as the
properties of dense astrophysical objects such as neutron stars.
Phenomenological or meson-theoretic two-body potentials, such as Argonne-V18 (AV18)
and CD-Bonn, were developed in the 1990s using constraints primarily from
nucleon-nucleon ($NN$) scattering
data~\cite{Machleidt:1987hj,Wiringa:1994wb}.  More recently, chiral
effective field theory (EFT) has led to the development of potentials
with systematic and controlled
approximations~\cite{Machleidt:2011,Epelbaum:2008ga}. 
Light atomic nuclei have played a crucial role in constraining modern
nuclear interaction models, including many-body forces, as many of
their properties (e.g., charge distributions and radii, ground- and
excited-state energies) can be both precisely measured and exactly
calculated for a given two- and three-nucleon interaction
model~\cite{Carlson:1997qn,Carlson:2014vla,Piarulli:2016,Hagen:2014,Barrett:2013,Lonardoni:2018prc}.

While the combination of $NN$ scattering and light-nuclei data allows
one to constrain the two- and three-nucleon interaction at large
distances, its short-ranged behavior is still largely
unconstrained. The latter is important for understanding
nucleon-nucleon short-range correlations (SRC) in nuclei
\cite{Hen:2016kwk,Atti:2015eda}, their relation to the partonic
structure of bound nucleons~\cite{Barak:Nature2019,Chen:2016bde,Weinstein:2010rt,Hen:2012fm,Hen:2013oha},
and the structure of neutron stars~\cite{Frankfurt:2008zv,Li:2018lpy}.

Constraining the short-ranged part of the nuclear interaction
requires studying nucleon momentum distributions at
high-momentum. However, previous attempts to extract these
were largely unsuccessful, due to the fact that nucleon
momentum distributions are not direct observables, and typical
experimental extractions suffer from large reaction mechanism effects.
These introduce significant model-dependent corrections that mask 
the underlying characteristics of the momentum distribution,
especially at
high-momentum~\cite{Bussiere:1981mv,Benmokhtar:2004fs,Rvachev:2004yr,Egiyan:2007qj}.

Advances in nuclear reaction theory now allow us to identify
observables with increased sensitivity to nucleon momentum densities
at
high-momentum~\cite{CiofidegliAtti:2005qt,Laget:2004sm,Alvioli:2009zy,Frankfurt:2008zv,Boeglin:2011mt}.
In light of these advances, we report on a new study of the momentum
distribution of nucleons in Helium-3 relative to Tritium over a broad
momentum range.

We study nucleon momentum distributions using Quasi-Elastic (QE)
electron scattering.  In these experiments, an electron with momentum
$\vec p_e$ is scattered from the nucleus, transferring energy $\omega$
and momentum $\vec q$ to the nucleus. We choose $\omega$ and $\vec q$
to be appropriate for elastic scattering from a moving bound nucleon.
By detecting the knocked-out proton ($\vec p_p$) in coincidence with
the scattered electron ($\vec p_e {\!'}$), we can measure the missing
energy and missing momentum of the reaction:
\begin{eqnarray}
E_{miss} &=& \omega - T_p - T_{A-1}, \\
\vec p_{miss} &=& \vec{p}_p - \vec{q},
\end{eqnarray}
where $\vec q = \vec p_e - \vec p_e {\!'}$ is the momentum transfer,
$T_{A-1} = (\omega + m_A - E_p) - \sqrt{(\omega + m_A - E_p)^2-|\vec{p}_{miss}|^2}$
is the reconstructed kinetic energy of the residual $A-1$ system, and
$T_p$ and $E_p$ are the measured kinetic and total energies of the outgoing proton.

% -------------------------------------------------------------------------------------------------------------
% Measurement and kinematics

In the Plane-Wave Impulse Approximation (PWIA) for QE scattering,
where a single exchanged photon is absorbed on a single proton and the
knocked-out proton does not re-interact as it leaves the nucleus, the
cross-section for \Aeep, electron-induced proton knockout from nucleus
$A$, can be written as~\cite{Kelly:1996hd,DeForest:1983ahx}:
\begin{equation}
\frac{d^6\sigma}{d\omega dE_p d\Omega_{e} d\Omega_{p}} = K \sigma_{ep} S(|\vec{p}_i|,E_i)
\label{eq.1}
\end{equation}
where $\sigma_{ep}$ is the cross-section for scattering an electron
from a bound proton \cite{DeForest:1983ahx}, $K=E_p |\vec{p}_p|$ is a
kinematical factor, $d\Omega_{e}$ and $d\Omega_{p}$ are the electron
and proton solid angles respectively, and $S(|\vec{p}_i|,E_i)$ is the
spectral function, which defines the probability to find a proton in
the nucleus with momentum $|\vec p_i|$ and separation energy $E_i$.
The nucleon momentum distribution is the integral of the spectral
function over the separation energy: $n(|\vec p_i|) =
\int{S(|\vec{p}_i|,E_i) dE_i} $.

In PWIA, the missing momentum and energy equal the initial momentum
and separation energy of the knocked-out nucleon: $\vec{p}_i =
\vec{p}_{miss}$, $E_i = E_{miss}$.  However, there are other, non-QE,
reaction mechanisms, including final state interactions (the
rescattering of the knocked-out proton, FSI), meson-exchange currents
(MEC), and exciting isobar configurations (IC) that can lead to the
same measured final state.  These also contribute to the cross
section, complicating this simple picture.  In addition, relativistic
effects can be significant \cite{gao00,udias99,AlvarezRodriguez:2010nb}.

Previous measurements of the \het\eep two- and three-body breakup
cross-sections were done at $Q^2 = 1.5$ (GeV/c)$^2$ and $x_B \equiv
\frac{Q^2}{2m_p\omega} = 1$ where $m_p$ is the proton
mass~\cite{Benmokhtar:2004fs, Rvachev:2004yr}, near the expected
maximum of the proton rescattering.  The measured cross-sections
disagreed by up to a factor of five with PWIA calculations for $\pmiss
> 250$ MeV/c.  These deviations were described to good accuracy by
calculations which included the contribution of non-QE reaction
mechanisms, primarily
FSI~\cite{CiofidegliAtti:2005qt,Laget:2004sm,Frankfurt:2008zv,Alvioli:2009zy}.
The large contribution of such non-QE reaction mechanisms to the
measured \eep cross-sections limited their ability to constrain the
nucleon momentum distribution at high momenta.

Guided by reaction mechanism calculations, which agree with previous
measurements, we can reduce the effect of FSI in two
ways~\cite{Boeglin:2011mt,Sargsian:2001ax,Frankfurt:1996xx,Jeschonnek:2008zg,Laget:2004sm,Sargsian:2009hf,Hen:2014gna}
by: (A) constraining the angle between
$\vec{p}_{recoil}=-\vec{p}_{miss}$ and $\vec{q}$ to be $\theta_{rq}
\lesssim 40^{\circ}$ and (B) taking the ratio of \eep cross-sections
for same-mass nuclei. The effect of FSI should be similar in both
nuclei because knocked-out protons in both nuclei can rescatter from
the same number of nucleons and FSI should therefore largely cancel in
the ratio.

Additional non-QE reaction mechanisms such as MEC and IC were shown to
be suppressed for $Q^2 \equiv q^2 - \omega^2 > 1.5$ (GeV/c)$^2$ and
$x_B > 1$~\cite{Sargsian:2001ax,Sargsian:2002wc}. Thus, the ratio of
\het\eep to \trit\eep cross-sections in QE kinematics at $Q^2> 1.5$
(GeV/c)$^2$, $x_B>1$ and $\theta_{rq} \lesssim 40^{\circ}$ should have
increased sensitivity to the ratio of their spectral functions.

We measured the ratios of $d$, \het, and \trit{} \eep cross-sections in Hall
A of the Thomas Jefferson National Accelerator Facility (JLab) using
the two high-resolution spectrometers (HRS) and a 20 $\mu A$ 4.326 GeV
electron beam incident on one of four 25-cm long gas target cells~\cite{target_report}.
The four identical cells were filled with Hydrogen ($70.8 \pm 0.4$
mg/cm$^2$), Deuterium ($142.2 \pm 0.8$ mg/cm$^2$), \het{} ($53.4 \pm
0.6$ mg/cm$^2$) and Tritium ($85.1 \pm 0.8$ mg/cm$^2$)
gas~\cite{Santiesteban:2018qwi}.  We detected the scattered electrons in the
left HRS at a central angle $\theta_e = 20.88^{\circ}$ and momentum
$p_e = 3.543$ GeV/c, corresponding to a central four-momentum transfer
$Q^2 = 2.0$ (GeV/c)$^2$, energy transfer $\omega = 0.78$ GeV, and
$x_{B} = 1.4$. We detected the knocked-out protons in the right HRS at
two different kinematical settings, $(\theta_p, p_p)$ =
($48.82^{\circ}$, 1.481 GeV/c), and ($58.50^{\circ}$, 1.246 GeV/c),
referred to here as ``low \pmiss'' and ``high \pmiss''
respectively. These two settings cover a combined missing momentum
range of $40 \le \pmiss \le 550$ MeV/c.
Deuterium measurements were only done in the ``low \pmiss'' kinematics
and thus extended only up to $\pmiss \sim 300$ MeV/c.

% ---------------------------------------
\begin{figure}
\includegraphics[scale=0.43]{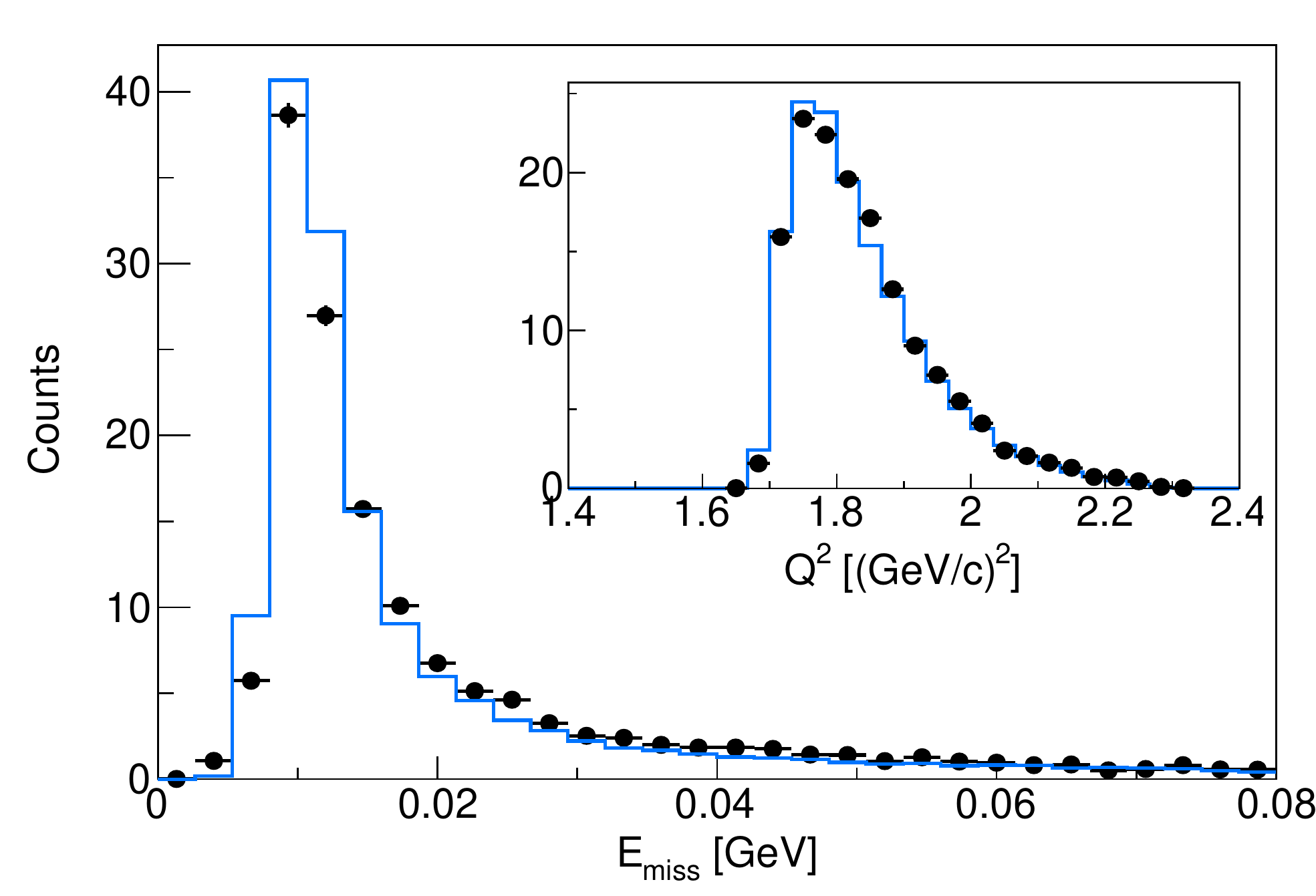}
\caption{(color online) Number of \trit\eep events (counts) versus
  missing energy for the low \pmiss{} kinematics.  The black markers
  correspond to the measured data. The lines correspond to the
  calculated distributions obtained from a SIMC \cite{Simc} simulation
  with a spectral function calculated by C. Ciofi degli Atti and
  L. P. Kaptari \cite{CiofidegliAtti:2004jg} and normalized to give
  the same integral as the
  data. Due to the lack of \trit{} proton spectral functions, we
  assumed isospin symmetry and used the \het{} neutron spectral
  function for the \trit\eep{} simulation. (see text for details).
  The insert shows the $Q^2$ distribution for the same kinematical
  setting. See online supplementary materials for equivalent \het{} distributions.}
\label{fig:kinematical}
\end{figure}
% ---------------------------------------

Each HRS consisted of three quadrupole magnets for focusing and one
dipole magnet for momentum analysis~\cite{Alcorn:2004sb,HRSupdates}.
These magnets were followed by a detector package, slightly updated
with respect to the one in Ref~\cite{Alcorn:2004sb}, consisting of a
pair of vertical drift chambers used for tracking, and two
scintillation counter planes that provide timing and trigger
signals. A CO$_{2}$ Cherenkov detector placed between the
scintillators and a lead-glass calorimeter placed after them were used
for particle identification.

% -------------------------------------------------------------------------------------------------------------
% Data analysis 

Electrons were selected by requiring that the particle deposits more
than half of its energy in the calorimeter: $\frac{E_{cal}}{|\vec{p}|}
> 0.5$. \eep coincidence events were selected by placing a
$\pm3\sigma$ cut around the relative electron and proton event times.
Due to the low experimental luminosity, the random coincidence event
rate was negligible.  We discarded a small number of runs with
anomalous numbers of events normalized to the beam charge.

Measured electrons were required to originate within the central $\pm 9$~cm
of the gas target to exclude events originating from the target walls.
The electron and proton reconstructed target vertices were required to be within 
$\pm 1.2$ cm of each other, which corresponds to $\pm3\sigma$ of the vertex 
reconstruction resolution.
By measuring scattering from an empty-cell-like target we determined that the 
target cell wall contribution to the measured \eep event yield was negligible ($\ll 1\%$).

To avoid the acceptance edges of the spectrometer, 
we restricted the analysis to events that are detected within $\pm 4\%$
of the central spectrometer momentum, 
and $\pm \SI{27.5}{\milli\radian}$ in in-plane angle and $\pm \SI{55.0}{\milli\radian}$
in out-of-plane angle relative to the center of the spectrometer acceptance.
In addition, we further restricted the measurement phase-space by 
requiring $\theta_{rq} < 37.5^{\circ}$ to minimize the effect of FSI and, 
in the high \pmiss{} kinematics, $x_B > 1.3$ to further suppress non-QE events.

The spectrometers were calibrated using sieve slit measurements to
define scattering angles and by measuring the kinematically over-constrained
exclusive H\eep and $^2$H$(e,e'p)n$ reactions. The H\eep reaction \pmiss{}
resolution was better than 9 MeV/c.
We verified the absolute luminosity normalization 
by comparing the measured elastic H\ee yield to a parametrization of the world data \cite{Lomon:2006xb}.
We also found excellent agreement between the elastic H\eep and H\ee rates,
confirming that the coincidence trigger performed efficiently.

Figure~\ref{fig:kinematical} shows the number of measured \trit\eep events as a
function of \emiss{} and of $Q^2$ for the low \pmiss{} setting as
well as the same distributions calculated using the Monte Carlo code
SIMC \cite{Simc} and normalized to give the same integrated number of
events as the data.
SIMC generated \eep events using Eq. (\ref{eq.1}), with the addition of radiation effects,
that were then propagated through the spectrometer model to account
for acceptance and resolution effects, and subsequently analyzed as the data.
The SIMC calculations used a \het{} spectral function calculated by C. Ciofi degli Atti 
and L. P. Kaptari using the AV18 potential \cite{CiofidegliAtti:2004jg}.
Due to the lack of \trit{} proton spectral functions, we assumed
isospin symmetry and used the \het{} neutron spectral function for the
\trit\eep simulation. The difference between the calculated momentum
distributions of neutrons in \hets and protons in \trits is small and
contributes a $3\%$ uncertainty to the $^3$H$(e,e'p)$ calculations and
to the spectral-function ratio calculations~\cite{Wiringa:2014}.  The
spectral function calculation appears to describe the measured $Q^2$
and \emiss{} distributions well. See online supplementary materials
for details and additional comparisons (including \het\eep{} spectra).

For each measured nucleus, we calculated the normalized \eep event yield as:
\begin{equation}
Y(\pmiss) = \frac{N(\pmiss)}{C \cdot t_{\rm{live}} \cdot (\rho/A) \cdot b},
\end{equation}
where $A$ is the target atomic weight, $N(\pmiss)$ is the number of counts
for that target in a given bin of \pmiss{} integrated over the experimental
\emiss{} acceptance, $C$ is the total accumulated beam charge,
$t_{\rm{live}}$ is the live time fraction in which the detectors are able to collect data,
$\rho$ is the nominal areal density of the
gas in the target cell, and $b$ is a correction factor to account for changes
in the target density caused by local beam heating. 
$b$ was determined by measuring the beam current dependence of
the inclusive event yield~\cite{Santiesteban:2018qwi}.  We formed
three yield ratios, \het/$d$, \trit/$d$, and \het/\trit.

We corrected the measured ratio of the normalized yields for the
radioactive decay of $2.78 \pm 0.18\%$ of the target \trits nuclei to
\hets in the six months since the target was filled, and denote the
corrected yield ratio by $R^{corr. yield}$.

% -------------------------------------------------------------------------------------------------------------
% Cross-section ratio extraction
The point-to-point systematical uncertainties on this ratio due to the
event selection criteria (momentum and angular acceptances, and
$\theta_{rq}$ and $x_B$ limits) were determined by repeating the
analysis 5000 times, selecting each criterion randomly within
reasonable limits for each iteration. The systematic uncertainty was
taken to be the standard deviation of the resulting distribution of
ratios.  They range from 1\% to 8\% and are typically much smaller
than the statistical uncertainties.  There is an overall normalization
uncertainty of $1.8\%$, predominantly due to the target density
uncertainty.  Other normalization uncertainties due to beam-charge
measurement and run-by-run stability are at the $1\%$ level or lower,
see Table~\ref{tab:uncertainties}.  See online supplementary materials
for details.

% ---------------------------------------
\begin{figure}
\includegraphics[scale=0.43]{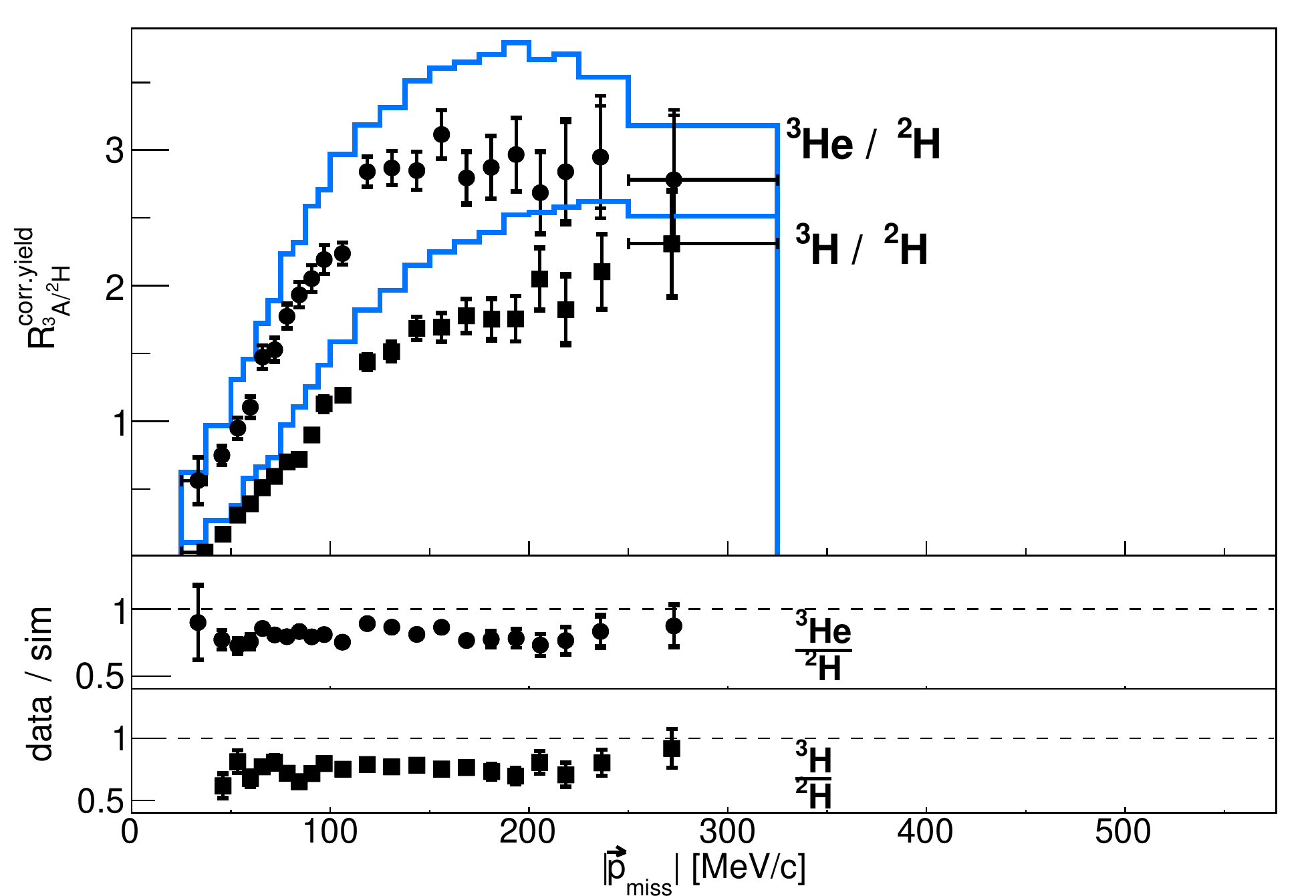}
\includegraphics[scale=0.43]{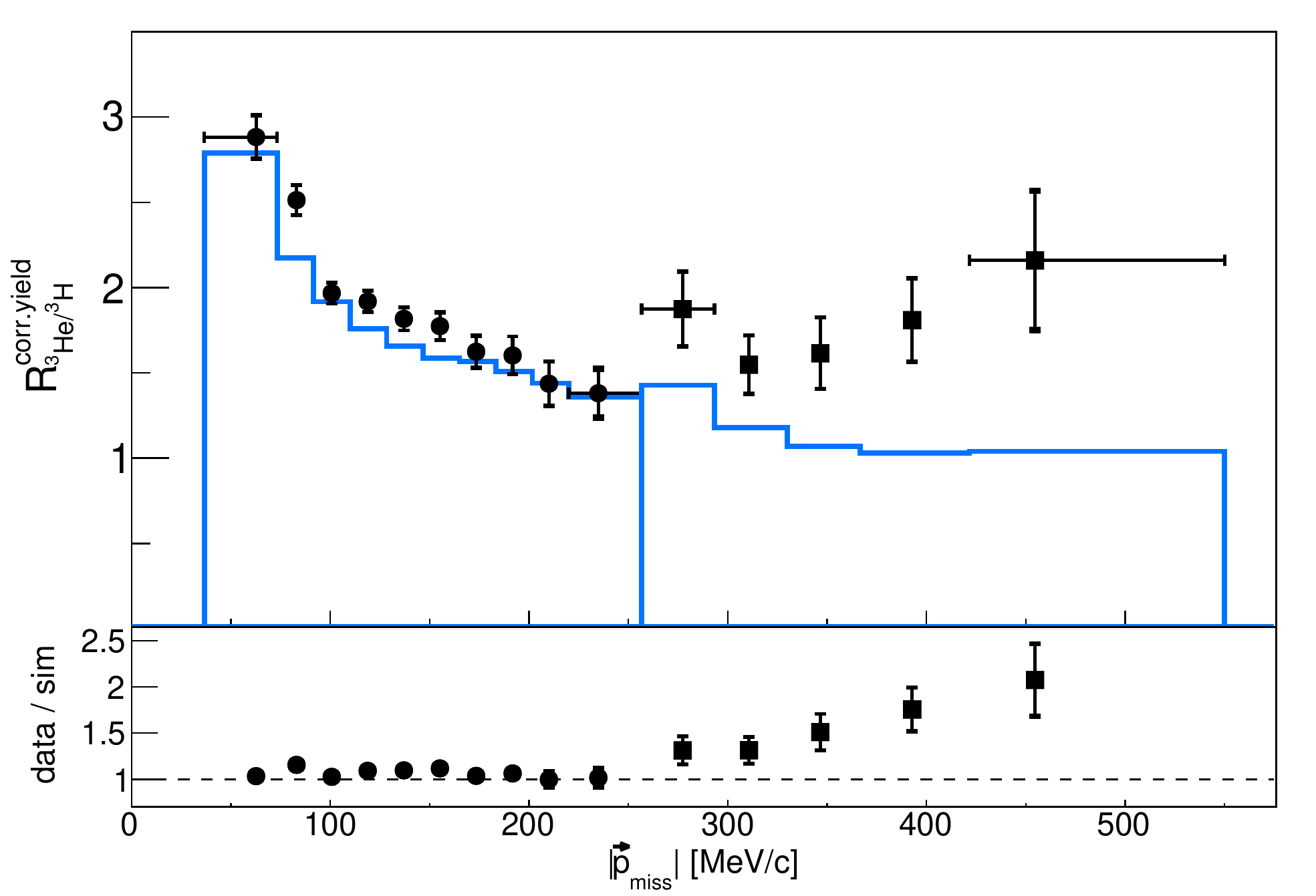}
\caption{Missing momentum dependence of the measured \eep \het/$d$ and
  \trit/$d$ (top) and \het/\trit (bottom) normalized event yield
  ratios.  The circles and squares correspond respectively to \het/$d$
  and \trit/$d$ in the top panel and to the low and high \pmiss{}
  settings in the bottom panel.  The error bars include both
  statistical and point-to-point systematical uncertainties.  An
  additional overall normalization uncertainty of $1.8\%$ is not shown
  (see Table~\ref{tab:uncertainties}).  The solid histogram shows the
  PWIA SIMC simulation using Eq. (\ref{eq.1}) and the spectral function of
  Ref.~\cite{CiofidegliAtti:2004jg} for $A=3$ and AV18 for $A=2$. The
  bin widths are the same for the histogram and the data.
}
\label{fig:3He_3H_uncor}
\end{figure}
% ---------------------------------------

Figure \ref{fig:3He_3H_uncor} shows the missing momentum dependence of
the corrected event yield ratios $R^{corr. yield}_{^3{\rm He} / d}$,
$R^{corr. yield}_{^3{\rm H} / d}$, and $R^{corr. yield}_{^3{\rm He} /
  ^3{\rm H}}$ for each kinematical setting.  The ratios of \het{} and
\trit{} to deuterium are very small at low \pmiss, due to the much
narrower deuterium momentum distribution, and increase to a constant value of
about two for \trit/$d$ and about three for \het/$d$ at the largest
measured \pmiss{} of about 270 MeV/c.  By contrast, the \het/\trit{}
ratio is about three at the smallest measured \pmiss{} and
decreases to about 1.5 at $\pmiss \approx 250$ MeV/c, with a possible
rise after that.  This is consistent with the low-\pmiss expectation
of 2.5 to 3 and slightly higher than the SRC-based high-\pmiss{}
expectation of one.  The change in the ratios is much smaller than the
four order-of-magnitude decrease in the calculated momentum
distributions (see online supplementary information).

Both measured \het/$d$ and \trit/$d$ ratios are about $20\%$ larger
than the PWIA spectral-function based SIMC calculation.  This
indicates that FSI effects are the same for both ratios.  For the same
missing momentum range, the measured and calculated \het/\trit{} ratios
agree within the measurement accuracy of about 3\%. This is a clear
indication for cancellation of FSI effect in the \het/\trit{} ratio.
At higher missing-momentum ($\pmiss > 250$ MeV/c), the measured
\het/\trit{} ratios are about $20 - 50\%$ larger than the calculation.

To extract the experimental cross-section ratio, $\sigma_{ \het\eep }
/ \sigma_{ \trit\eep } (p_{miss})$, we corrected the measured yield
ratios using SIMC for radiative and
bin-migration effects as well as for the finite \emiss{}
acceptance of the spectrometers.  The finite \emiss{} correction equals the calculated
momentum distribution ratio divided by the calculated ratio of
spectral functions integrated over the missing energy acceptance.
The individual and total corrections were all less
than 10\% for all $\pmiss$ values.  We apply a
point-to-point systematic uncertainty of 20\% of the resulting
correction factors.  See Table~\ref{tab:uncertainties} and online
supplementary material for details.

We also calculated the final state interaction effects of single
rescattering of the knocked-out proton with either of the two other
nucleons in the three-body-breakup reaction in the generalized Eikonal
approximation \cite{misak05a,misak05b} using a computer code developed
by M. Sargsian~\cite{SargsPRivate}.  For each bin we calculated both
the PWIA and FSI cross section and integrated over the experimental
acceptance.  FSI changed the individual \het{} and \trit\eep{}
cross-sections by between 10\% and 30\%.  However, they largely
cancelled in the double ratio
\begin{equation}
R^{FSI} =
\frac{\sigma_{FSI}/\sigma_{PWIA}\vert_{\het}}{\sigma_{FSI}/\sigma_{PWIA}\vert_{\trit}}.
\end{equation}
producing at most a 5\% effect at the highest \pmiss. This reinforces
the claim that FSI effects are very small in the cross-section ratio.
We did not correct the data for FSI.  See online supplementary
materials for more information.

We tested the cross section factorization approximation
by comparing the factorized spectral function approach used in SIMC
with an unfactorized calculation by J. Golak
\cite{CARASCO200341,BERMUTH2003199,Golak:2005iy}.  The difference
between the factorized and non-factorized calculations was about
$5\%$, which is not enough to explain the data-calculation discrepancy
at high \pmiss.

% -------------------------------------------------------------------------------------------------------------
% Results
% ---------------------------------------
\begin{figure}
\includegraphics[scale=0.43]{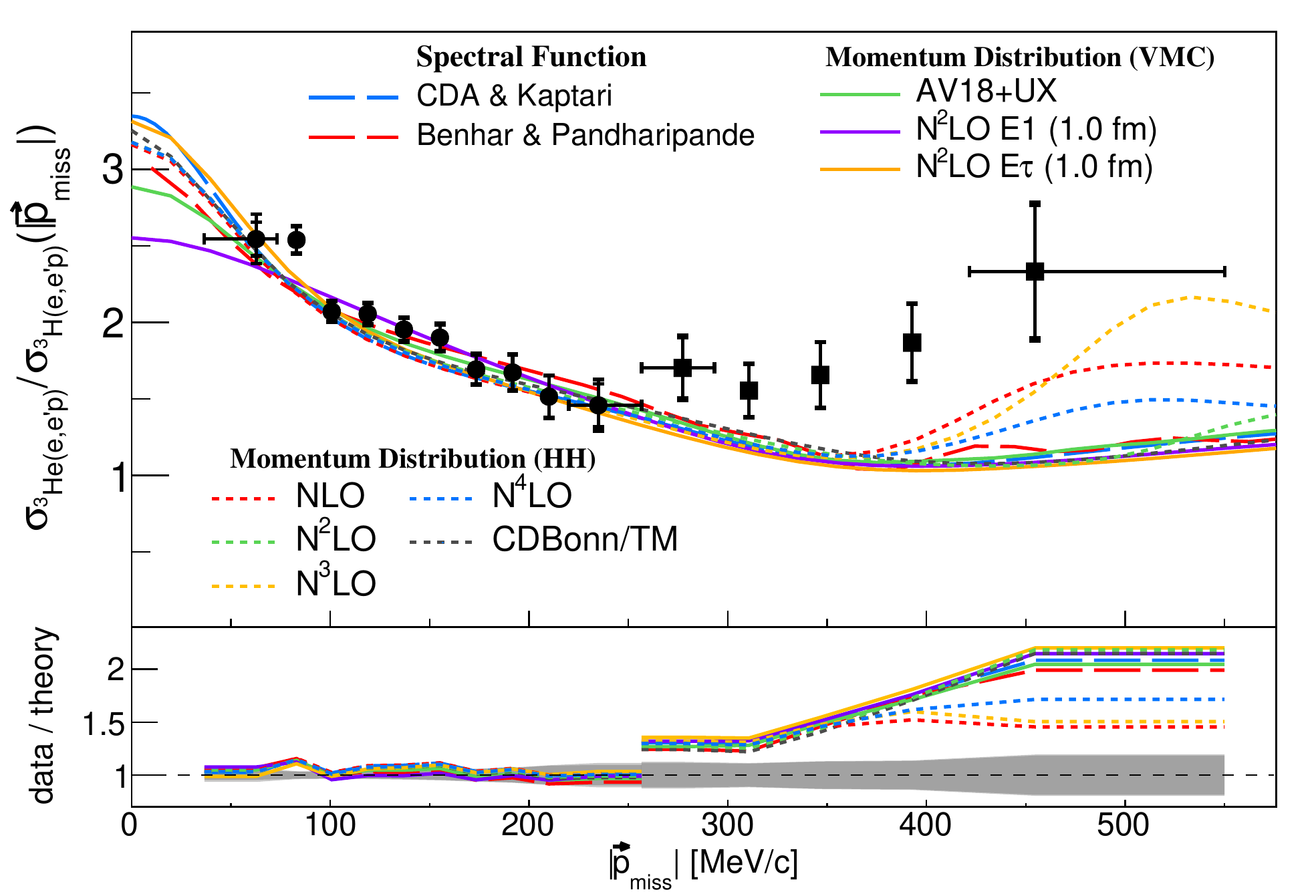}
\caption{(color online) 
  The measured \het{} to \trit{} cross-section ratio,  $\sigma_{\het\eep} / \sigma_{\trit\eep}  (p_{miss})$, 
  plotted vs. \pmiss{} compared with different models of
  the corresponding momentum distribution ratio.  The filled circle and
  square markers correspond to the low and high \pmiss{} settings
  respectively. Uncertainties shown include both statistical and
  point-to-point systematical uncertainties.  The 
  overall normalization uncertainty of about $1.8\%$ is not shown (see table~\ref{tab:uncertainties}).
  Horizontal bars indicate the bin sizes and are shown
  for only the first and last points in each kinematical setting as
  all other points are equally spaced.  The bottom panel shows the
  double ratio of data to different calculated momentum distribution ratios,
  with the grey band showing the data uncertainty.
  The theoretical calculations are done using different local and non-local interactions,
  as well as different techniques for solving the three-body problem. See text for details.}
\label{fig:3He_3Hfinal}
\end{figure}
% ---------------------------------------

Figure \ref{fig:3He_3Hfinal} shows the \pmiss{} dependence of the
extracted \het/\trits \eep{} cross-section ratio.  In the simplest model, this ratio should
equal two, the relative number of protons in \het{} and \trit.
However, at large \pmiss{} the ratio should equal one,
the relative number of $np$ SRC pairs in \het{} and
\trit~\cite{Weiss:2016obx,piasetzky06,tang03,shneor07,subedi08,korover14,Hen:2014nza,Duer:2018sby,Duer:2018sxh}.
These SRC pairs will shift equal amounts of cross-section strength
from low \pmiss{} to high \pmiss{} in both nuclei, increasing the
\het{} to \trit{} ratio at low \pmiss{} to more than two.  The
measured ratio follows this simple model of a transition from
independent nucleons at the lowest \pmiss{} to $np$-SRC pairs at
higher \pmiss, decreasing
from almost three at low \pmiss{} towards about 1.5 at $\pmiss = 250$ MeV/c.
At larger \pmiss{} the measured ratio is approximately flat, with a
possible rise at the largest \pmiss. 

With the missing-energy acceptance correction for \het/\trit{} and the
small expected FSI effects, the resulting cross-section ratios should
be sensitive to the ratio of momentum distributions.  We therefore
compare in Fig.~\ref{fig:3He_3Hfinal} the measured cross-section
ratios directly with the ratio of various single-nucleon momentum
distributions. The momentum distribution calculations are obtained
using either the variational Monte Carlo (VMC) technique with local
interactions~\cite{Wiringa:2014,Lonardoni:2018nofk} or the
Hyperspherical Harmonics (HH)
method~\cite{Kievsky:2008es,Marcucci:2018llz} with non-local
interactions.
    
The local interactions used include the phenomenological
AV18~\cite{Wiringa:1994wb} two-nucleon potential augmented by the
Urbana X (UX)~\cite{Wiringa:nofk} three-nucleon force and the chiral
EFT potentials at N$^2$LO (including two- and three-body
contributions), using a coordinate-space cutoff of $1$ fm and
different parametrizations of the three-body contact term $E\tau$ and
$E1$~\cite{Gezerlis:2014,Lynn:2016,Lynn:2017,Lonardoni:2018prl,
  Lonardoni:2018prc}.  Non-local interactions include the
meson-theoretic CD-Bonn~\cite{Machleidt:2000ge} two-nucleon potential,
together with the Tucson-Melbourne~\cite{Coon:2001pv} (TM)
three-nucleon potential, or the latest chiral two-body potentials from
NLO to N$^4$LO \cite{Entem:2017gor}, including three-nucleon
interactions. The main contribution to the latter, namely the one
arising from two-pion exchange, is effectively included at the same
chiral order as the two-nucleon interaction, as explained in
Refs.~\cite{Entem:2017gor,Marcucci:2018llz}. In these calculations, the
momentum-space cutoff $\Lambda$ is kept fixed at 500 MeV.  The VMC
calculations using the AV18 and UX interactions produce equivalent
results as the HH calculations using the AV18 plus Urbana
IX~\cite{Pudliner:1995wk} interactions.
    
For completeness, Fig.~\ref{fig:3He_3Hfinal} also shows the
momentum-distribution ratio calculated by integrating over the missing
energy in the spectral functions of Ref.~\cite{CiofidegliAtti:2004jg}
and Ref.~\cite{Benhar:1993ja}, obtained using the AV18 two-nucleon
only and the AV14~\cite{Wiringa:1984tg} two- and the Urbana
VIII~\cite{Carlson:1983kq} (UVIII) three-nucleon interactions,
respectively.
    
All calculated momentum-distribution ratios shown agree with the data
up to $\pmiss \approx 250$ MeV/c. At larger \pmiss, the theoretical
predictions obtained by integrating the spectral functions or by
calculating the momentum distribution ratio with local potentials or
with the CD-Bonn/TM model disagree with the data by 20--50\%. In the
case of the non-local chiral potential models, the calculations show
significant order dependence.
    
Note that, while momentum distributions calculated with local
chiral-interactions depend strongly on the cutoff parameter, these
effects appear to mostly cancel in the ratio of the momentum
distributions~\cite{Lonardoni:2018sqo}.
    
Finally, although FSI calculated in the generalized Eikonal approximation are
small, more complete calculations are needed, including two-
and three-body interaction operators \cite{more17}, to determine if
the discrepancy between data and calculation is due to the reaction
mechanism or to the validity of the underlying $NN$ potentials at
short-distances.  In addition, fully relativistic calculations are
needed to see if there are any significant corrections due to
longitudinal-transverse interference effects
\cite{gao00,udias99,AlvarezRodriguez:2010nb}.

One possible explanation for the discrepancy could be single-charge
exchange FSI, where a struck neutron from an SRC rescatters at almost
180$^\circ$ from a proton, and the proton is detected ($np$ SCX), or a
struck proton from an SRC rescatters at almost 180$^\circ$ from a
neutron ($pn$ SCX).  A struck proton in an SRC rescattering from its
partner neutron will decrease the number of observed proton events and
a struck neutron in an SRC rescattering from its partner proton will
increase the number of observed proton events.  These two effects will
largely cancel in both \het\eep{} and \trit$(e,e'p)$.  However, in \het{}
the struck neutron in an SRC can rescatter from the uncorrelated
proton, increasing the number of observed proton events but in \trit{}
it cannot.  This can increase the observed \het/\trit{} ratio.  In
addition, if the SCX occurs at $\theta < 180^\circ$, then events at
small \pmiss{} will be observed at larger \pmiss, amplifying the
effects of SCX at large \pmiss.

% ---------------------------------------
\begin{table}[]
  \caption {Systematic uncertainties in the extraction of the \het/$d$,
    \trit/$d$, and \het/\trits \eep normalized event-yield ratios,
    $R^{corr. yield}_{^3{\rm He} / ^3{\rm H}}$,
    (Fig.~\ref{fig:3He_3H_uncor}) and the \het/\trits cross-section
    ratio, $\sigma_{\het\eep} / \sigma_{\trit\eep}$,
    (Fig.~\ref{fig:3He_3Hfinal}). Uncertainties marked by `*'
    contribute only to the cross-section ratio. All uncertainties are
    summed in quadrature. See text for details.}
\label{tab:uncertainties}
\begin{tabular}{cccll}
\cline{1-3}
\multicolumn{1}{|c|}{}                                                                                                                  & \multicolumn{1}{c|}{Overall}                   & \multicolumn{1}{c|}{Point-to-point} &  &  \\ \cline{1-3}
\multicolumn{1}{|c|}{Target Walls}                                                                                                      & \multicolumn{1}{c|}{$\ll 1$\%} & \multicolumn{1}{c|}{}               &  &  \\ \cline{1-3}
\multicolumn{1}{|c|}{Target Density}                                                                                                    & \multicolumn{1}{c|}{1.5\%}                     & \multicolumn{1}{c|}{}               &  &  \\ \cline{1-3}
\multicolumn{1}{|c|}{Beam-Charge and Stability}                                                                                         & \multicolumn{1}{c|}{1\%}                       & \multicolumn{1}{c|}{}               &  &  \\ \cline{1-3}
\multicolumn{1}{|c|}{Tritium Decay}                                                                                                     & \multicolumn{1}{c|}{0.18\%}                     & \multicolumn{1}{c|}{}               &  &  \\ \cline{1-3}
\multicolumn{1}{|c|}{Cut sensitivity}                                                                                                   & \multicolumn{1}{c|}{}                          & \multicolumn{1}{c|}{1\% - 8\%}      &  &  \\ \cline{1-3}
\multicolumn{1}{|c|}{\begin{tabular}[c]{@{}c@{}}Simulation Corrections*\\ (bin-migration, radiation,\\ $E_{m}$ acceptance)\end{tabular}} & \multicolumn{1}{c|}{}                          & \multicolumn{1}{c|}{1\% - 2\%}      &  &  \\ \cline{1-3}
\multicolumn{1}{l}{}                                                                                                                    & \multicolumn{1}{l}{}                           & \multicolumn{1}{l}{}                &  & 
\end{tabular}
\end{table}
% ---------------------------------------

% -------------------------------------------------------------------------------------------------------------
% Discussion
To summarize, we presented the first simultaneous measurement of the
\het$(e,e'p)$, \trit\eep and $d$\eep reactions in kinematics where the
cross-sections are expected to be sensitive to the proton momentum
distribution, i.e., at large $Q^2$, $x_B>1$, and
$\theta_{rq}<40^\circ$ that minimize two-body currents and the effects
of FSI.  We further enhanced the sensitivity to the momentum
distribution by extracting the ratio of the cross-sections, so that
most of the remaining FSI effects cancel, as confirmed by a
generalized Eikonal approximation calculation of leading proton
rescattering.

The measured \het/$d$ and \trit/$d$ corrected yield ratios are small at low
\pmiss{} and increase to three and two respectively at $\pmiss = 250$
MeV/c.   Both are about 20\% lower than PWIA calculated yield ratios, indicating that FSI effects are about
the same in both pairs of reactions.

While the measured corrected cross-section ratio $\sigma_{\het\eep} /
\sigma_{\trit\eep}$ is well described by PWIA calculations up to
$\pmiss\approx 250$ MeV/c, they disagree by only 20 - 50\% at high \pmiss,
despite a four order of magnitude decrease of the
momentum distribution in this range (see Fig.~2 of the online
supplementary information).  
This is a vast improvement over previous $\sigma_{\het\eep}$ measurements
at lower $Q^2$ and $x_B = 1$, which disagreed with PWIA calculations
by factors of several at large \pmiss~\cite{Benmokhtar:2004fs,
  Rvachev:2004yr}.   This, together with FSI calculations, strongly
supports the reduced contribution of non-QE reaction mechanisms in our
kinematics. 

The data overall supports the transition from single-nucleon dominance
at low \pmiss, towards an $np$-SRC pair dominant region at high
\pmiss~\cite{Weiss:2016obx,piasetzky06,tang03,shneor07,subedi08,korover14,Duer:2018sby,Duer:2018sxh,Hen:2014nza}.
However, more complete calculations are needed to assess the
implications of the observed 20--50\% deviation of the data from the PWIA
calculation in the expected $np$-SRC pair dominance region, including
the effects of single charge exchange.  If the observed
 difference between the \het/\trit{} experimental ratio and momentum
distribution ratios at large missing momenta is due to the underlying
$NN$ interaction, then it can provide significant new constraints on
the previously loosely-constrained short-distance parts of the $NN$
interaction.

% ==============================================================================================================
% Acknowledgement
We acknowledge the contribution of the Jefferson-Lab target group and
technical staff for design and construction of the Tritium target and
their support running this experiment. We thank C. Ciofi degli Atti
and L. Kaptari for the $^3$He spectral function calculations and
M. Sargsian, M. Strikman, J.~Carlson, S.~Gandolfi, and R.~B.~Wiringa
for many valuable discussions. This work was supported by the
U.S. Department of Energy (DOE) grant DE-AC05-06OR23177 under which
Jefferson Science Associates, LLC, operates the Thomas Jefferson
National Accelerator Facility, the U.S. National Science Foundation,
the Pazi foundation, the Israel Science Foundation, and the NUCLEI
SciDAC program.  Computational resources for the calculation of the
N$^2$LO momentum distributions have been provided by Los Alamos Open
Supercomputing via the Institutional Computing (IC) program and by the
National Energy Research Scientific Computing Center (NERSC), which is
supported by the U.S. Department of Energy, Office of Science, under
Contract No. DE-AC02-05CH11231. The Kent State University contribution
is supported under the PHY-1714809 grant from the U.S. National Science Foundation.
The University of Tennessee contribution is supported by the DE-SC0013615 grant.
The work of ANL group members is supported by DOE
grant DE-AC02-06CH11357.

% ==============================================================================================================
% Bibliography
\bibliography{TritiumBib}

\end{document}